\documentclass[%
reprint,
showpacs,
amsmath,amssymb,
aps,
pra,
]{revtex4-1}
\usepackage{graphicx}% Include figure files
\usepackage{dcolumn}% Align table columns on decimal point
\usepackage{bm}% bold math
\usepackage{hyperref}% add hypertext capabilities
\usepackage[mathlines]{lineno}% Enable numbering of text and display math
\usepackage[usenames]{color}
\usepackage{dsfont}
\usepackage{amssymb}
\usepackage[T1]{fontenc}
\usepackage{enumerate}
\usepackage{fancyhdr}
\usepackage{eqnarray,amsmath}\usepackage[usenames]{color}
\usepackage{relsize}
\usepackage{epstopdf}
\usepackage{subfigure}
\usepackage{float}
\usepackage{verbatim}
\usepackage{blkarray}

\begin{document}

\title{Hexapartite entanglement in an above-threshold Optical Parametric Oscillator}

\author{
F. A. S. Barbosa\textsuperscript{1},
A. S. Coelho\textsuperscript{2,3},
L. F. Mu\~n{}oz-Mart\'i{}nez\textsuperscript{4},
L. Ortiz-Guti\'e{}rrez\textsuperscript{5},
A. S. Villar\textsuperscript{6},
P. Nussenzveig\textsuperscript{7},
M. Martinelli\textsuperscript{7*}
}

\affiliation{
\textsuperscript{1}  Instituto de F\'\i{}sica Gleb Wataghin, Universidade Estadual de Campinas, 13083-859 Campinas, SP, Brazil\\
\textsuperscript{2} Dept. de Engenharia Mec\^anica, Universidade Federal do Piau\'\i{}, 64049-550 Teresina, PI, Brazil. \\
\textsuperscript{3} Dept. de Engenharia, Centro Universit\'ario UNINOVAFAPI, 64073-505 Teresina, PI, Brazil \\
\textsuperscript{4} Dept. de Ciencias B\'a{}sicas, Universidad del Sin\'u{}-El\'\i{}as Bechara Zain\'u{}m, Cra 1w \# 38-153, Monter\'\i{}a, C\'o{}rdoba, Colombia\\
%\textsuperscript{4} Departamento de F\'\i{}sica, Universidade Federal de Pernambuco, 50670-901 Recife, PE, Brazil  \\
\textsuperscript{5} Instituto de F\'\i{}sica de S\~a{}o Carlos, Universidade de S\~a{}o Paulo, P. O. Box 369, 13560-970 S\~a{}o Carlos, SP, Brazil\\
\textsuperscript{6} American Physical Society, 1 Research Road, Ridge, New York 11961, USA \\
\textsuperscript{7} Instituto de F\'\i{}sica da Universidade de S\~a{}o Paulo, P.O.Box 66318, 05315-970 S\~a{}o Paulo, Brazil
}

\email{mmartine@if.usp.br} %% email address is required

\begin{abstract}
%We investigate, both theoretically and experimentally, the structure of hexapartite entanglement among sideband modes of an optical parametric oscillator operating above threshold. This widely used system, in CW operation, generates a rich structure of multimode entanglement, which can be tuned by a single parameter, the pump power. Furthermore, given the difference in wavelengths of pump, signal, and idler carrier modes, entanglement can be used to controllably distribute quantum information throughout the electromagnetic spectrum.
%PN

We demonstrate, theoretically and experimentally, the generation of hexapartite modal entanglement by the optical parametric oscillator (OPO) operating above the oscillation threshold. We show that the OPO generates a rich structure of entanglement among sets of six optical sideband modes interacting through the non-linear crystal. The class of quantum states thus produced can be controlled by a single parameter, the power of the external laser that pumps the system. Our platform allows for the generation of massive entanglement among many
%tens of thousands of
optical modes  with well defined but vastly different frequencies, potentially bridging nodes of a multicolor quantum network.

\end{abstract}

\maketitle

In the burgeoning field of quantum information science~\cite{NielsenChuang}, entanglement is considered to be the greatest resource. This intrinsic quantum property, studied since the early days of quantum mechanics~{\cite{EPR,Bohr}, can be generated in a number of physical systems and particularly in quantum optics, owing to the great control over optical systems and the high fidelity in the measurement of theirs observables.

One of the workhorses of the field, the continuously pumped triply resonant optical parametric oscillator (CW OPO), consists of a nonlinear crystal that couples three modes within a cavity (Fig. \ref{opo}). The nonlinear coupling leads to the creation (and annihilation) of pairs of photons in downconverted fields (denoted 1 and 2, also known as signal and idler), with the annihilation (or creation) of a photon in the pump field 0. Since the pumped nonlinear crystal acts as a gain medium, when this gain matches the cavity losses, the system achieves an oscillatory regime with the generation of intense output beams. Controlling the pump power, we can explore a broad set of different quantum states of the field. Examples range from squeezed states for the downconverted mode \cite{kimblesqz} and the pump \cite{pumpsqz}, to bipartite entanglement below \cite{pengentangled} and above \cite{villarentangled} the oscillation threshold, reaching tripartite entanglement involving fields spanning more than one octave in frequency \cite{tripartite}.

Further steps are typically required to generate more intricate multipartite entangled states. For instance,  off-cavity combinations of squeezed states and beam splitters lead to a two-rail cluster state generation in the time domain, presenting entanglement of more than 10,000 modes defined by multiplexing of a CW OPO output in time slices of 160 ns \cite{Furusawa}. Alternatively, by manipulating the pairwise generation of entangled states in frequency modes separated by the cavity free spectral range,  quadripartite entangled states \cite{Pfister}, and a frequency comb of 60 modes separated by 1 GHz \cite{Pfister2} were engineered. Multipartite entanglement was also generated with pulsed OPOs, leading to entanglement over the wide spectra of its output, as studied in 10 spectral modes in the range from 790 to 800 nm \cite{Fabre_multi}.

\begin{figure}[h]
  \begin{center}
    \includegraphics[width= 1.0\linewidth]{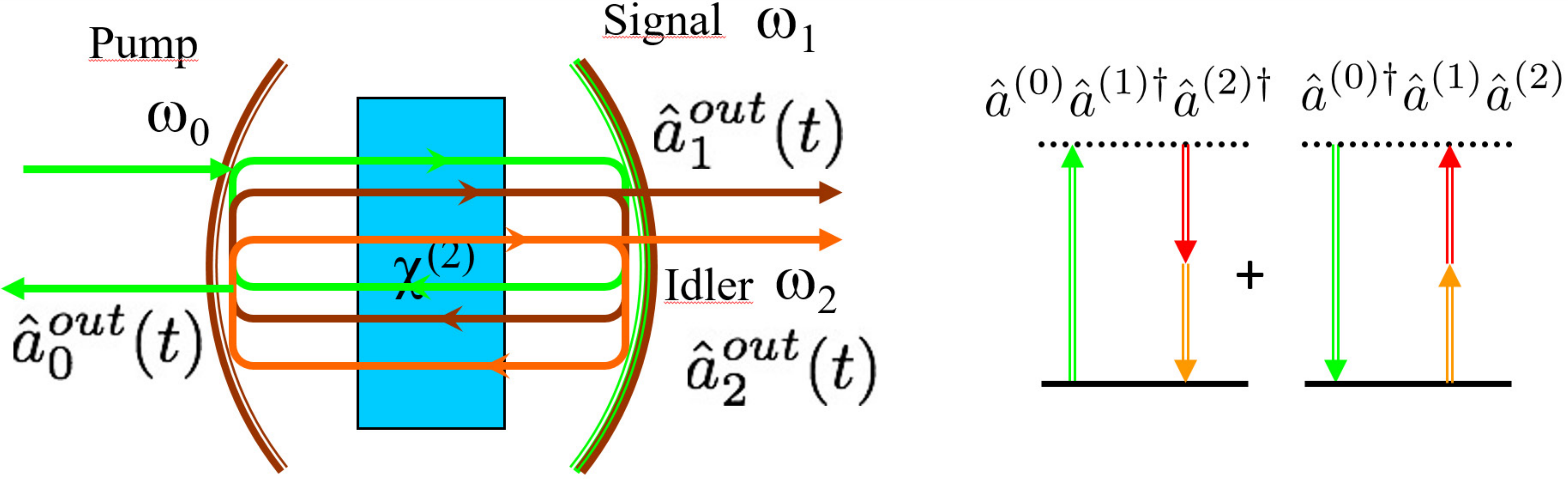}
    \caption{The optical parametric oscillator consists of one triply resonant cavity with a nonlinear crystal, that is responsible for coupling the pump field (0) to the signal (1) and idler (2) downconverted fields. }
    \label{opo}
  \end{center}
\end{figure}

In this Letter, we show that, even without resorting to such techniques, a rich structure of multimode entangled states is already found in the CW triply resonant OPO
pumped by a monochromatic field, in its operation above the oscillation threshold (Fig. \ref{opo}).
Starting from a driving field, with vacuum states for all the other modes, the system evolves to an hexapartite entangled state.
This entanglement is found in the sideband modes of the intense fields of the reflected pump and the downconverted beams.
These modes are accessed by a combination of electronic demodulation of the measured photocurrents
of the fields, with the help of empty cavities for each beam in a resonator detection technique \cite{hexaopo}. Hexapartite entanglement is verified by tests of positivity under partial transposition~\cite{Peres_1996}.
In the following, we present the structure of the generated entangled states as well as their control by the pump power.

%Considering the whole bandwidth covered by these sidebands, a massive set of hexapartite entangled modes could be generated, and by multifrequencial pump, these systems can be coupled in a rich multipartite entangled structure.

The evolution of the field operators $\hat a^{(n)}$ for each resonant cavity mode depends on the propagation inside the cavity, the coupling to external modes through the mirrors, and their coupling within the nonlinear crystal. The latter is described by an interaction Hamiltonian
\begin{eqnarray}\label{a1}
\hat{H}_{\chi}=i\hslash \dfrac{\chi}{\tau}\left[ \hat{a}^{(0)}(t)\hat{a}^{(1)\dagger}(t)
\hat{a}^{(2)\dagger}(t)-\text{h.c.}\right],
\end{eqnarray}
where $\chi$ is the effective second order susceptibility,  $\tau$ is the time of flight through the medium and  field indices 0, 1 and 2 stand for pump, signal and idler modes, respectively. The OPO has been studied in detail for decades \cite{reiddrummond}, and it is usually treated by the time evolution of these field operators $\hat a^{(n)}$, or by the evolution of the density operator in a suitable quasi-probability representation. These treatments lead to an effective three-mode description of the problem.

Nevertheless, a careful analysis of the measurement technique \cite{prlsideband} reveals that the measured state of the output modes involves information of the two sideband modes for each carrier field. The role of each individual sideband is clear if we consider that each annihilation operator of the field $\hat{a}^{(n)}(t)$ is associated  to the electric field operator of a propagating wave that can be described by the sum of time independent field operators  as
$\hat{a}^{(n)}(t)=e^{-i\omega_{n}t}\int_{-\omega_{n}}^{\infty} d\Omega e^{-i\Omega t} \hat{a}_{\omega_{n}+\Omega}^{(n)}$, where $\hat{a}_{\omega}$ %^{(n)}\equiv \hat{a}^{(n)}(\omega_{n}+\Omega)$
is the photon annihilation operator in the mode of frequency $\omega=\omega_{n}+\Omega$, and the
carrier frequency $\omega_n$ is put in evidence. On the detection of the output fields, we access the information on the sideband modes, shifted by $\Omega$ from the carrier \cite{hexaopo,prlsideband}.

The interaction Hamiltonian can be rewritten using a linearized version of the field operators in the rotating frame, detailing the role of the sideband modes. In this linearized description, the field operator is replaced by its mean value $\alpha_{\omega_{n}}=\langle \hat{a}^{(n)}(t)e^{i\omega_{n}t}\rangle = \langle \hat{a}_{\omega_{n}}^{(n)} \rangle$ and a fluctuation term
 $\delta \hat{a}^{(n)}(t)= \hat{a}^{(n)}(t)e^{i\omega_{n}t}-\alpha_{\omega_{n}}$.
If we retain only the terms satisfying the rotating wave approximation, and neglect those without the contribution of the intense field amplitudes $\alpha_{\omega_{n}}$, we have the Hamiltonian  involving the specific sideband modes of the three carriers, with  $\Omega>0$,
\begin{multline}
\hat{H}_{\chi}(\Omega)=-i\hslash \dfrac{\chi}{\tau}\Big[  \alpha_{\omega_{0}}^{*}\Big( \hat{a}^{(1)}_{\omega_{1}+\Omega} \hat{a}^{(2)}_{\omega_{2}-\Omega}+
\hat{a}^{(1)}_{\omega_{1}-\Omega}\hat{a}^{(2)}_{\omega_{2}+\Omega}\Big)+
\\
 \alpha_{\omega_{1}}\left(
  \hat{a}^{(0)\dagger}_{\omega_{0}+\Omega}\hat{a}^{(2)}_{\omega_{2}+\Omega}
  +\hat{a}^{(0)\dagger}_{\omega_{0}-\Omega}\hat{a}^{(2)}_{\omega_{2}-\Omega}\right)+
 \\
 \alpha_{\omega_{2}}\left(
 \hat{a}^{(0)\dagger}_{\omega_{0}+\Omega} \hat{a}^{(1)}_{\omega_{1}+\Omega}
 +\hat{a}^{(0)\dagger}_{\omega_{0}-\Omega}\hat{a}^{(1)}_{\omega_{1}-\Omega}\right) -\text{h.c.} \Big],
\label{a6}
\end{multline}
where we have discarded the constant term $(\alpha_{\omega _{0}}^{*}\alpha_{\omega_{1}}\alpha_{\omega_{2}}-\text{c.c.})$, which will just introduce a global phase.
Linear terms like $ \alpha_{\omega_{0}}^{*}\alpha_{\omega_{1}}\hat{a}^{(2)}_{\omega_{2}-\Omega}$ were also discarded because they lead to phase space displacement, which doesn't change entanglement properties. Moreover, they will typically average to zero, since they don't satisfy phase matching.
%where we have discarded the term $(\alpha_{\omega _{0}}^{*}\alpha_{\omega_{1}}\alpha_{\omega_{2}}-\text{c.c.})$, which will return a constant value and no contribution to the dynamics of the system.
%Linear terms like $ \alpha_{\omega_{0}}^{*}\alpha_{\omega_{1}}\hat{a}^{(2)}_{\omega_{2}-\Omega}$ will include an oscillatory contribution $\exp(-i\Omega t)$ that doesn't satisfy energy conservation.
The total Hamiltonian is given by the sum of the contributions for each positive frequency $\Omega$, as $\hat{H}_{\chi}=\int_{\epsilon}^{\infty} \hat{H}_{\chi}(\Omega) \, d\Omega$. Therefore, the detailed treatment of the state of the sideband modes associated with a single analysis frequency $\Omega$ is decoupled from those of frequencies $\Omega '\neq\Omega$ \cite{Munos}.
%, as far as third order terms of like $  \hat{a}^{(0)\dagger}_{\omega_{0}+\Omega} \hat{a}^{(1)}_{\omega_{1}+\Omega'} \hat{a}^{(2)}_{\omega_{2}-\Omega''}$.

\begin{figure}[h]
  \begin{center}
    \includegraphics[width= 0.65\linewidth]{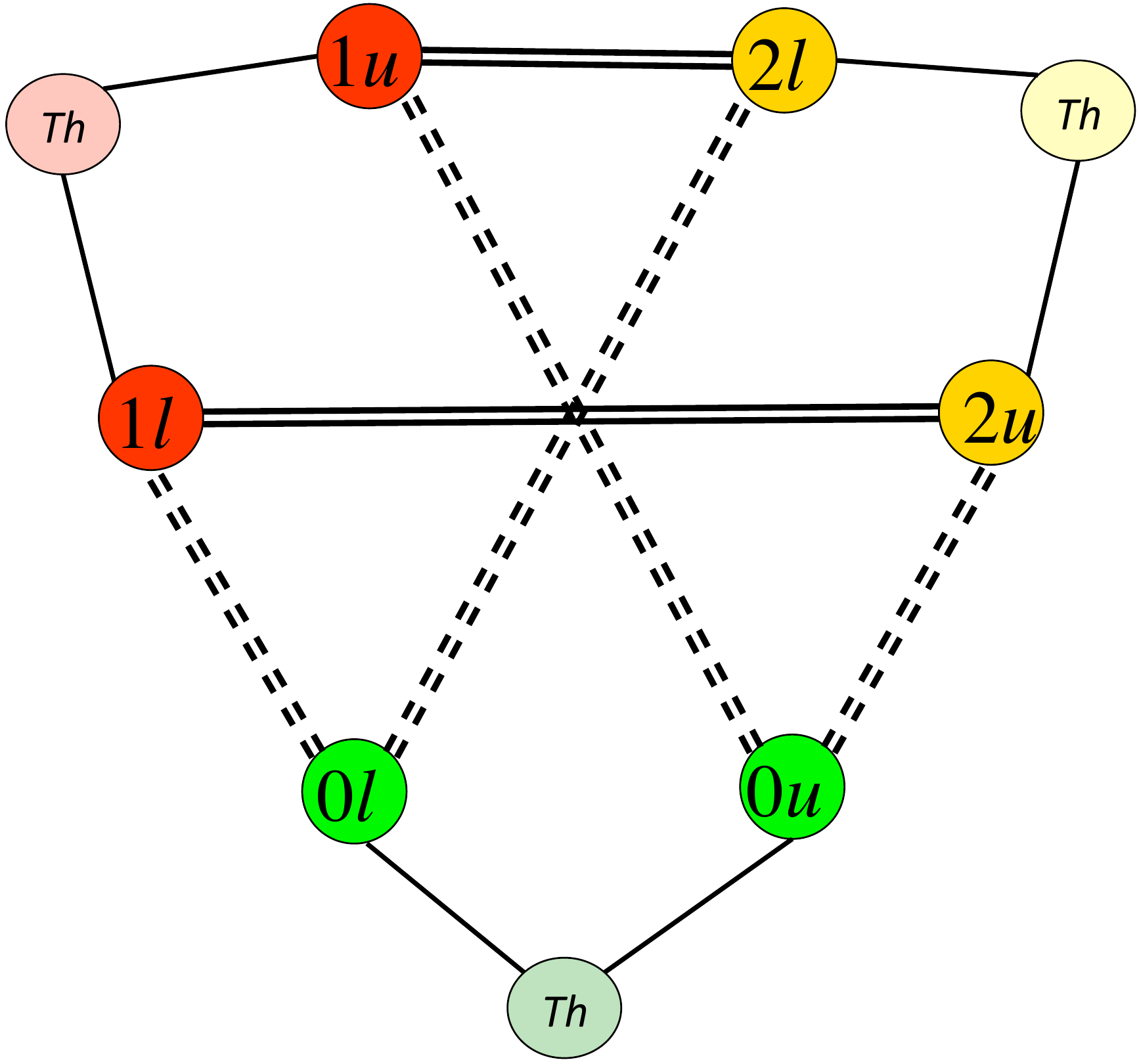}
    \caption{Coupling of the six sideband modes of the field. Signal and idler sidebands are coupled by photon pair creation (and annihilation) operators (double lines). All the other modes are pairwise coupled by beam-splitter operations (double dashed lines). Each sideband pair is coupled to thermal reservoirs (Th) by phonon scattering (straight single lines). Notation: $iu$ ($il$) stands for upper (lower) sideband modes at frequency $\omega_i+\Omega$ ($\omega_i-\Omega$).}
    \label{hexa}
  \end{center}
\end{figure}

%The resulting Hamiltonian includes, therefore, a pair of terms involving the creation and annihilation of pairs of photons in symmetric sideband modes of signal and idler fields, mediated by the mean pump field. This term is associated with two-mode squeezing operations, and it will be the only leading term when the OPO is below threshold, resulting in entangled EPR states \cite{pengentangled} or squeezing in the case of degenerate signal and idler modes \cite{kimblesqz}. Above threshold, mean field amplitudes for signal and idler are non-zero, and the other four terms will imply in photon exchange between the pump sidebands and the downconverted sidebands, mediated by the other downconverted mean field. These beam splitter operations will couple all the six sidebands in a ring, leading to a cascaded coupling  among all the modes. The result is an hexapartite entangled state, controlled by the mean fields (Fig. \ref{hexa}).

If the evolution of the system could be described just by the unitary operations in Eq. (\ref{a6}) and  the cavity dynamics without spurious losses, the resulting state would be pure, with  entanglement for each one of the 31 possible bipartitions in the above threshold operation.
Nevertheless, Brillouin scattering of carrier photons by phonons of the crystal should be taken into account for intense intracavity fields \cite{phononpra}. From the optomechanical Hamiltonian \cite{optomecHamilt}, an extra contribution to the Hamiltonian of the form below can be derived
\begin{eqnarray}
 \hat{H}_{g}(\Omega)=\sum_{n=0}^{2}\sum_{j=1}^{3}-\hslash g_{nj}\left[ \alpha_{\omega_{n}}\left( \hat{a}_{\omega_{n}-\Omega}^{(n)\dag}\hat{d}_{\Omega}^{(j)\dag}+ \right. \right.\nonumber\\ \left.\left.
\hat{a}_{\omega_{n}+\Omega}^{(n)\dag}\hat{d}_{\Omega}^{(j)} \right)+ \text{h.c.}\right], \label{Eq:pH3}
\end{eqnarray}
with $\hat{d}^{(j)}_{\Omega_{m}}$ as the phonon annihilation operator on the mode of  frequency $\Omega_{m}$ in longitudinal and transverse mechanical modes indicated by index $j$. It will couple the sideband modes to different thermal reservoirs of the crystal, thus degrading purity and entanglement even for perfect cavities.
This phonon coupling appears to be intrinsic to the system, but it can, in principle, be mitigated  by cooling down the crystal \cite{tripartite}.
%, however other techniques could be used. For instance, stimulated Brillouin scattering could be can be mitigated in optical fibers by decreasing the spatial overlap between the optical and the acoustic modes and doping \cite{https://ieeexplore.ieee.org/document/1636371/citations?tabFilter=papers}.

Therefore, the state of the sidebands depends on the mean fields and it can be directly related to the normalized pump power $\sigma$ \cite{Debuisschert93} for exact resonance, taking the oscillation threshold as $\sigma=1$. Variation of this single parameter enables the exploration of this rich structure of nonclassical fields. Moreover, since only bilinear terms are involved, the resulting state is Gaussian, as experimentally observed in \cite{gaussian}.

Entanglement is directly generated by the two mode squeezing operator, associated to the creation and annihilation of photon pairs in different modes, presented in Eq. (\ref{a6}). This term is the only one remaining in sub-threshold operation, leading to squeezed states or entanglement \cite{kimblesqz,pengentangled}, coupling separate pairs of sidebands (Fig. \ref{hexa}). Above threshold, as the mean field of the downconverted modes grows with the increasing pump power, the coupling of the entangled modes to sidebands of the pump by the beam splitter operators will transfer information to these modes, and couple the formerly independent pair of entangled states. The system would remain pure, but the coupling to the phonon modes, and the loss of information in their reservoirs, degrades information in the system as the power grows, as can be seen from Eq.  (\ref{Eq:pH3})

In what follows, we study the OPO described in \cite{Munos}, using a KTP crystal inside a cavity with a free spectral range of 4.3(5) GHz and finesses of 15 for the pump mode, and 125 for signal and idler modes. Transmittance of cavity mirrors is 30 \% for the pump and 4 \% for the infrared couplers.
We performed a complete measurement of the covariance matrix
of the Hermitian quadrature operators $(\hat p, \hat q)$ of the field, associated to the annihilation operator $\hat a=\hat p +i\hat q$, for all the six modes involved, with overall quantum efficiencies of 65\% for the pump and  87\% for the infrared \cite{sm}.
Entanglement in this system is observed by the analysis of the physicality of the  smallest symplectic eigenvalue $\tilde\nu$ of the  covariance matrix for a partially transposed density operator of the state \cite{Simon_2000}.  Whenever $\tilde\nu< 1$, there is entanglement between the bipartitions \cite{sm}.
Experimental results are presented in Figs. (\ref{fig3},\ref{fig4}), confronted with the calculated values derived from our complete model of the OPO \cite{Munos}
(straight lines) and those without the phonon coupling (dashed lines). We have selected five representative cases from the complete set of results, which can be found in the Supplemental Material.

\begin{figure}[h]
  \begin{center}
    \includegraphics[width= \linewidth]{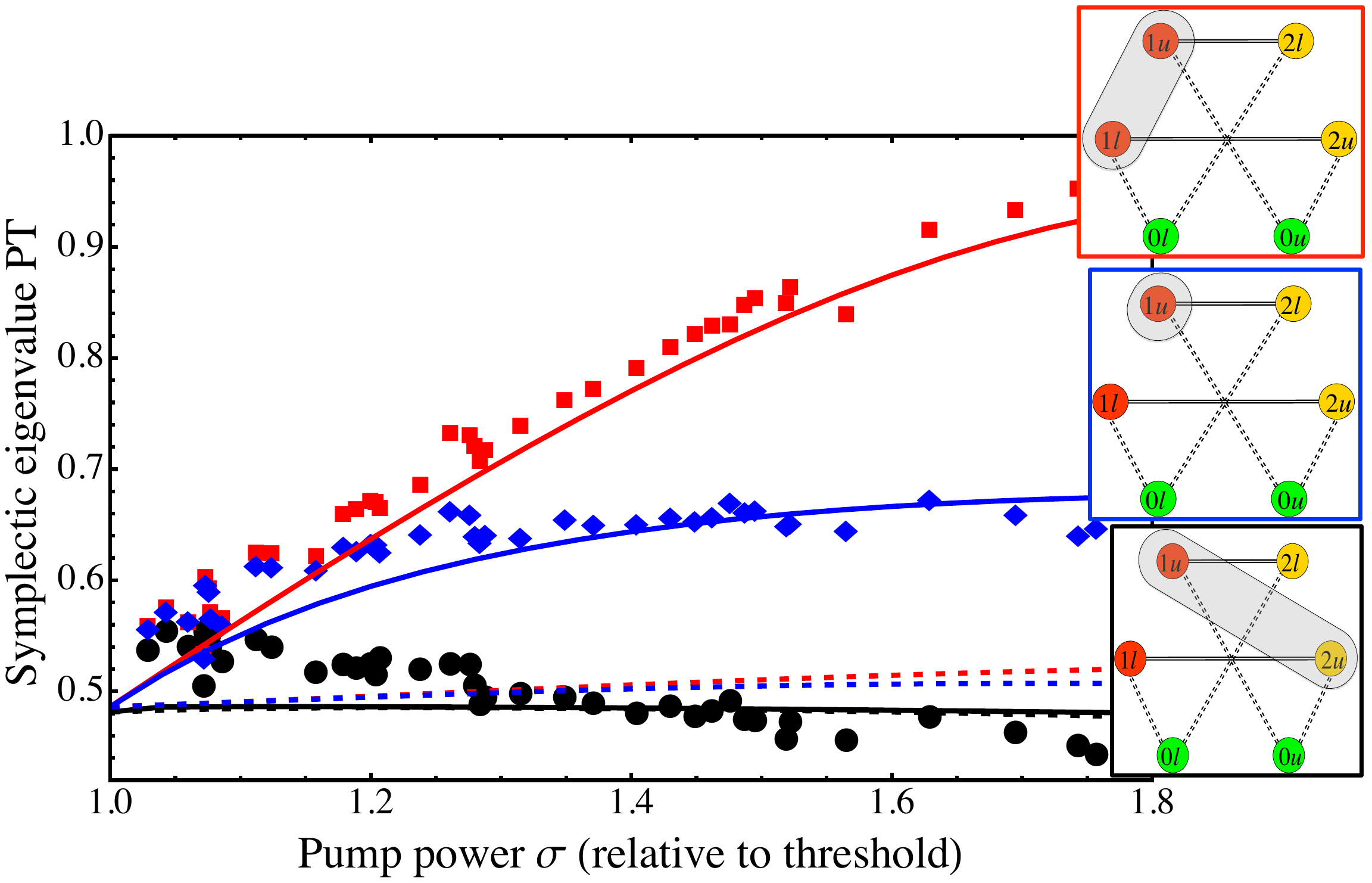}
        \caption{Symplectic eigenvalues from the transpositions of different bipartitions. Experimental results are compared with the complete model (straight lines) and the model without phonons (dashed). In the inset, the modes selected for one partition are marked by the gray shadow.}
    \label{fig3}
  \end{center}
    \end{figure}

%MM
We begin the analysis by those bipartitions where the originally entangled modes lie in separate bipartitions, in Fig. (\ref{fig3}). We can take  both upper sidebands of signal and idler in one partition (black), both sidebands of the same field in one partition (red), or select a single sideband in one partition (blue). In all these situations, in the absence of phonons, the symplectic eigenvalue remains nearly unchanged for growing pump powers. A small change is observed in the second and third case as the power increases, indicating a transfer of information to the sidebands for the pump. Nevertheless, if we include those sidebands in the selected partition \cite{sm}, deviations from this behavior are small.
%MM
%

The situation changes dramatically in  the presence of phonons. The growing coupling of the entangled modes to thermal reservoirs degrades the entanglement for some bipartitions, eventually leading to separability \cite{tripartite}. It is interesting to notice that taking pairs of sidebands of different beams in the same partition provides a protected configuration (Fig. \ref{fig3}, black). This situation resembles the robustness of the twin beam squeezing \cite{twinfabre}, originated from the parity in the photon creation in the downconverted modes.
On the other hand, if both sidebands from the same mode are taken in the same partition, phonon scattering leads to fast degradation of entanglement (Fig. \ref{fig3}, red). The difference between these cases can be understood from the fact that the coupling of each field to a thermal reservoir implies in correlated noise injected on the sidebands of that particular field, as a random phase modulation of the central carrier \cite{phononpra}. If we consider bipartition of the kind $1u2u\times1l2l$, additional noise in mode $1u$ is correlated to the noise added in mode $1l$, therefore information in both partitions remains correlated. The same applies to modes $2u$ and $2l$. On the other hand, for bipartitions of the kind $1l1u\times2l2u$, fluctuations in mode 1 are not perfectly correlated to those of mode 2 and this additional noise degrades the overall correlation, leading to a reduction on the observed entanglement. It is curious to notice that if we take just one of the modes in the partition (Fig. \ref{fig3}, blue), we have an intermediate situation, since we are comparing it to a set of modes where just one of them remains strongly correlated.

\begin{figure}[h]
  \begin{center}
\includegraphics[width= \linewidth]{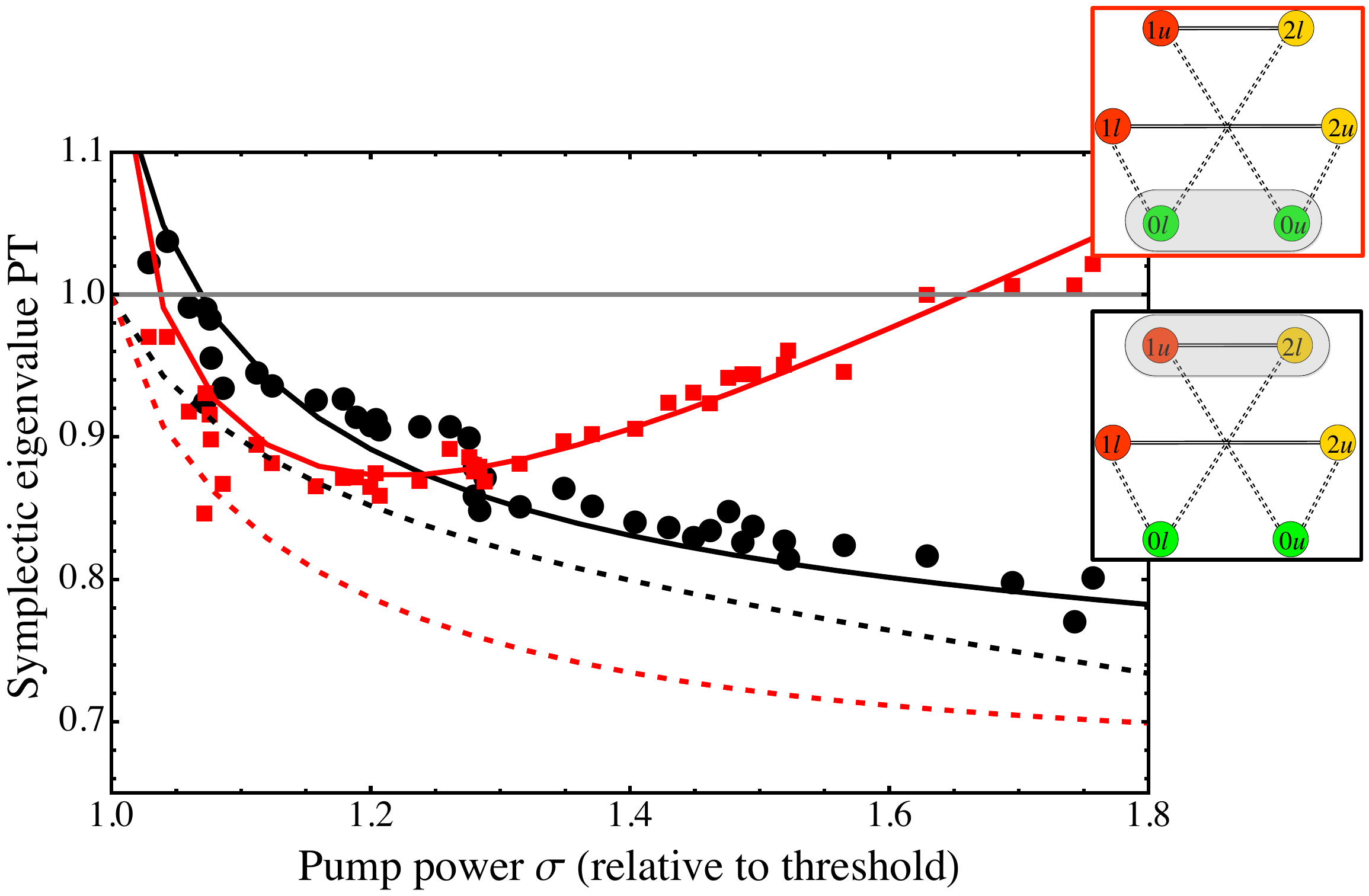}
    \caption{Symplectic eigenvalues from the transpositions of different bipartitions. Experimental results are compared with the complete model (straight lines) and the model without phonons (dashed). In the inset, the modes selected for one partition are marked by the gray shadow.}
    \label{fig4}
  \end{center}
\end{figure}

The role of the pump sidebands and their coupling to the entangled pairs remained almost unnoticed in the cases studied in Fig. \ref{fig3}. But they have an important effect in other bipartitions for growing pump powers, as can be seen in Fig. \ref{fig4}. If we separate each pair of entangled modes in different partitions (Fig. \ref{fig4}, black), we have completely independent sets while operating below the threshold, and therefore no entanglement between them. Above threshold, the signal and idler intensities are increased by growing pump power, and these pairs of modes are coupled to the pump sidebands (as can be seen in Eq. \ref{a6}), that intermediates the exchange of information between these two sets. The result is a growing entanglement between them, that can be improved if the sidebands of the pump are evenly distributed between the bipartitions \cite{sm}. In the absence of phonon scattering, the result would be further improved, since the downconverted modes would be coupled to the same reservoir (the pump sidebands), whose evolution we can follow from the measurement of the pump.

The pump modes become  necessarily entangled upon interaction with the signal and idler modes. In Fig. \ref{fig4}, the red curve shows the entanglement witness for the partition involving both pump sidebands. The observed entanglement is consistent with the one observed for an efective three-mode description of the OPO \cite{tripartite}. Once again, since the pump modes are coupled to the thermal reservoir of phonons, it suffers from the uncorrelated noise that is added to the sidebands and, as seen in the case of the single beam partition (Fig. \ref{fig3} red), it is strongly affected in the case of intense intracavity fields, which grows with the pump power. The result is quite similar if we take only one of the sidebands in a $1 \,\times \,5$ partition, as can be seen in the Supplemental Material.

%MM
The modes under study are selected by the choice of the demodulation frequency in the detection process. They are characterized by the analysis frequency, limited by the OPO cavity bandwidth (in the present case, in the range of 34 MHz), and the bandwidth of the detection, or the measurement rate in the acquisition system (600 kHz). Different modes can be accessed just by changing the analysis frequency. All these independent hexapartite systems are simultaneously generated by a single monochromatic pump field. The present configuration of our system, detailed in the Supplemental Material, provides at least 20 sets of hexapartite entangled modes, each set independent from the other.

%MM

On the other hand, we can expect that, if multi-frequency pump fields are used, with a frequency separation smaller than the cavity bandwidth, each intense pump mode can be treated classically by its mean value, and their fluctuations will now be coupled in the cavity in a situation similar to that shown in ~\cite{Pfister2}. It opens the path to generate massive multipartite entanglement in this system from the multiple coupling of these hexapartite entangled states. Comparing to the other approaches~\cite{Furusawa,Pfister2,Fabre_multi}, a similar large number of modes can be entangled. Here, we have the benefit of entangling sideband modes of carriers with very different wavelengths, enabling the distribution of quantum information over much broader, albeit discontinuous, bandwidths.

In conclusion, we have fully analyzed and characterized entanglement among six sideband modes in a triply resonant above-threshold OPO. The choice of the sidebands under study is done by the choice of the analysis frequency in the detected photocurrent.
 The rich structure of the entanglement generated can be easily controlled by means of the pump power, providing important flexibility for this platform. A controllable and scalable source of entangled states for quantum information tasks can thus be envisioned. The role of phonon scattering was also investigated, indicating that, by temperature control of the crystal, entanglement and purity of the system can be improved.

%\end{widetext}
%\end{comment}
%\section*{Acknowledgements}

The authors acknowledge  support from grant \# 2010/08448-2, \href{http://dx.doi.org/10.13039/501100001807}{Funda\c c\~ao de Amparo \`a Pesquisa do Estado de S\~ao Paulo (FAPESP)}, project 473847/2012-4  by the \href{http://dx.doi.org/10.13039/501100003593}{Conselho Nacional de Desenvolvimento Cient\'\i fico e Tecnol\'ogico},  by the Instituto Nacional de Ci\^encia e Tecnologia de Informa\c{c}\~ao Qu\^antica (INCT-IQ),
and by Coordena\c c\~ao de Aperfei\c coamento de Pessoal de N\'\i vel Superior.
%MM
M. Martinelli would like to thank Stephen P. Walborn for the discussions about the paper.
The authors would like to thank the anonymous referees for fruitful suggestions that greatly contributed to a better presentation of the results.
%MM

%\bibliography{Draft_Hexapartite}% Produces the bibliography via BibTeX.

%%%%%%%%%%%%%%%%%%%%%%%%%%%%%%%%%%%%%%%%%%%%%%%%%%%%%%%%%%%%%%%%%%%%%%%%%%%%%%%%%%%%%%%%%%%%%%%%%%%%%%%%%%%%%%%%%%%%%%%%%%%%%%

\clearpage

\newpage

\section{Supplemental Material}

\subsection{Experimental Setup}

The cavity of the Optical Parametric Oscillator consists of an input coupler (IC) allowing the injection of the pump field, and an output coupler, as the output port for the downconverted fields, as can be seen in Fig.\ref{fig:setup}. These mirrors are made of concave substrates with curvature radius of 50 mm. For the input coupler, the deposited reflective coating has a reflectivity of 70$\%$ at 532 nm and high reflectivity ($>99\%$) at wavelengths close to 1064 nm. The infrared output coupler (OC) has a reflectivity of 96$\%$ at $\approx$1064 nm and high reflectivity ($>99\%$) at 532 nm. Flat surfaces are AR coated for the specific wavelengths.
The crystal is a type II phase-matched KTP (potassium titanyl phosphate, KTiOPO$_{4}$) with length $l=12$ mm, average refractive index n=1.81(1) and antireflective coatings for both wavelengths.
Given the length of the cavity, the average free spectral range for the three modes is found to be of 4.3(5) GHz. The cavity finesse for pump mode is 15 and 124 for the signal and idler modes (the latter defined as the mode with the same polarization as the pump).

\begin{figure}[h]
  \begin{center}
    \includegraphics[width=\columnwidth]{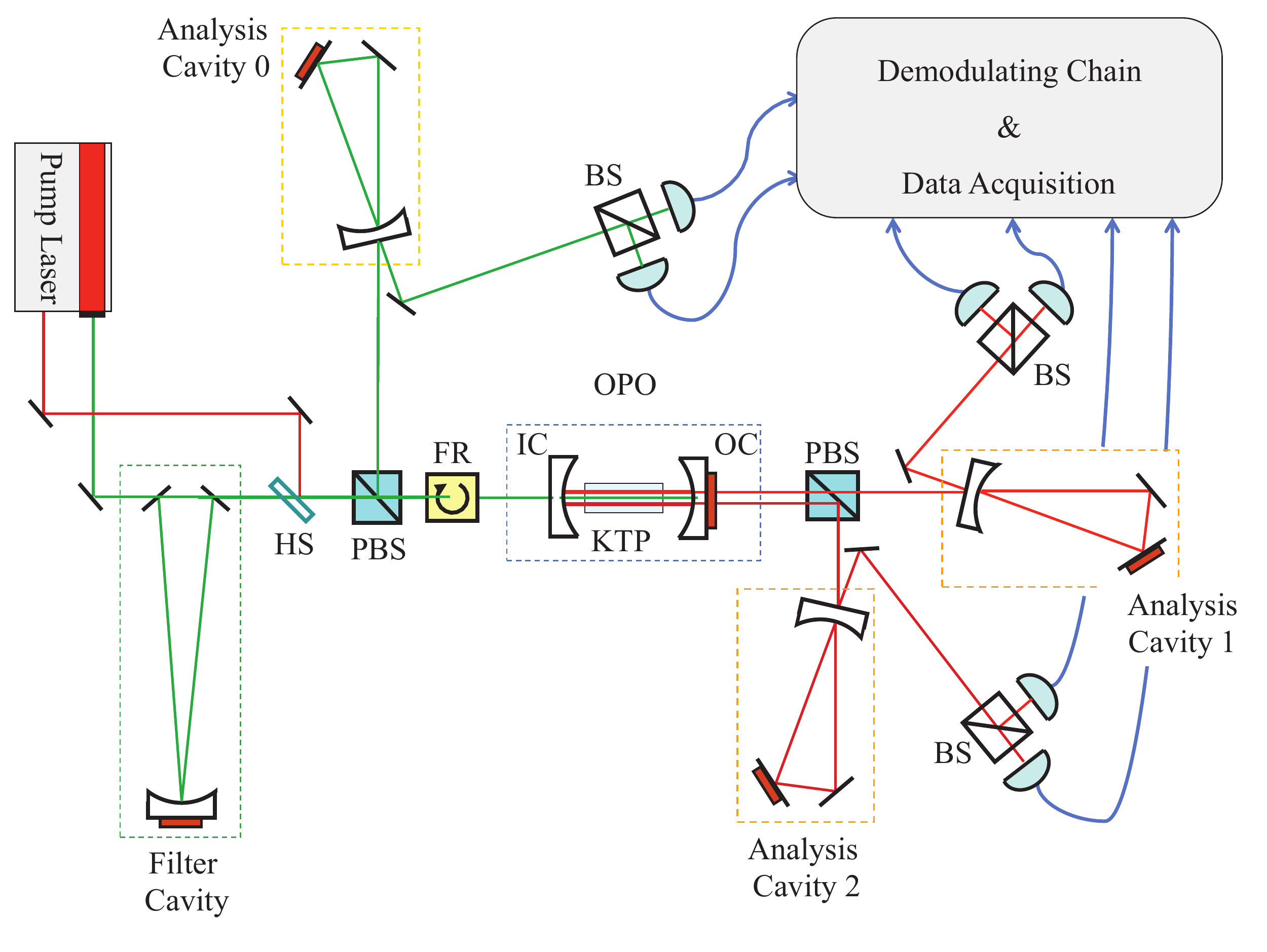}
    \caption{Setup for the reconstruction of the OPO beams' covariance matrix, as presented in \cite{hexaopo}. PBS, polarizing beam splitter; BS, 50:50 beam splitter; HS, harmonic separator; IC, input coupler; OC, output coupler (OPO cavity); FR, Faraday rotator.}
    \label{fig:setup}
  \end{center}
\end{figure}

%The OPO is pumped by the second harmonic of a doubled Nd:YAG laser (Innolight's Diabolo).  To ensure the low noise level of the pump field, the beam is filtered by a mode cleaning cavity, ensuring that pump fluctuations are reduced to the standard quantum level in amplitude and phase for frequencies above 20 MHz.
The monochromatic pump field generated by a doubled Nd:YAG laser (Innolight's Diabolo) has a linewidth of 1kHz and it is spectrally filtered by an optical resonator with a bandwidth of 800 kHz,  ensuring that pump fluctuations are reduced to the standard quantum level in amplitude and phase for frequencies above 20 MHz.
The filtered pump beam is then injected in the OPO, and the power can be adjusted by the polarization of the field, transmitted by a polarizing beam splitter (PBS).  The reflected pumped field is recovered from the same beam splitter, since the field undergo a double passage through a Faraday rotator (FR) after the reflection of the the cavity. The output fields close to 1064 nm are separated by a polarizing beam splitter. The original beam at the Nd:YAG wavelength, provided by the manufacturer, is used for alignment and test of the whole setup, and is injected in the OPO through the IC with the help of an harmonic separator (HS).
The threshold power is 60 mW, and the maximum pump power was 75\% above the threshold. In order to reduce the effect of phonon noise on the system, the crystal is cooled to 260 K, and the OPO is kept in a  vacuum chamber to avoid condensation.

Phase noise measurements were performed using the ellipse rotation method described in \cite{Galatola1991,Villar2008}, with the help of analysis cavities. Cavities 1 and 2 (for the transmitted infrared beams) have bandwidths of 14(1) MHz, and cavity 0 (for the reflected pump) has a bandwidth of 12(1) MHz. This ensures a full rotation of the noise ellipse
for the chosen analysis frequency of 21 MHz. Mode matching of the beams to the analysis cavities was better than 95 $\%$.

\begin{figure}[h]
  \begin{center}
    \includegraphics[width=0.7\columnwidth]{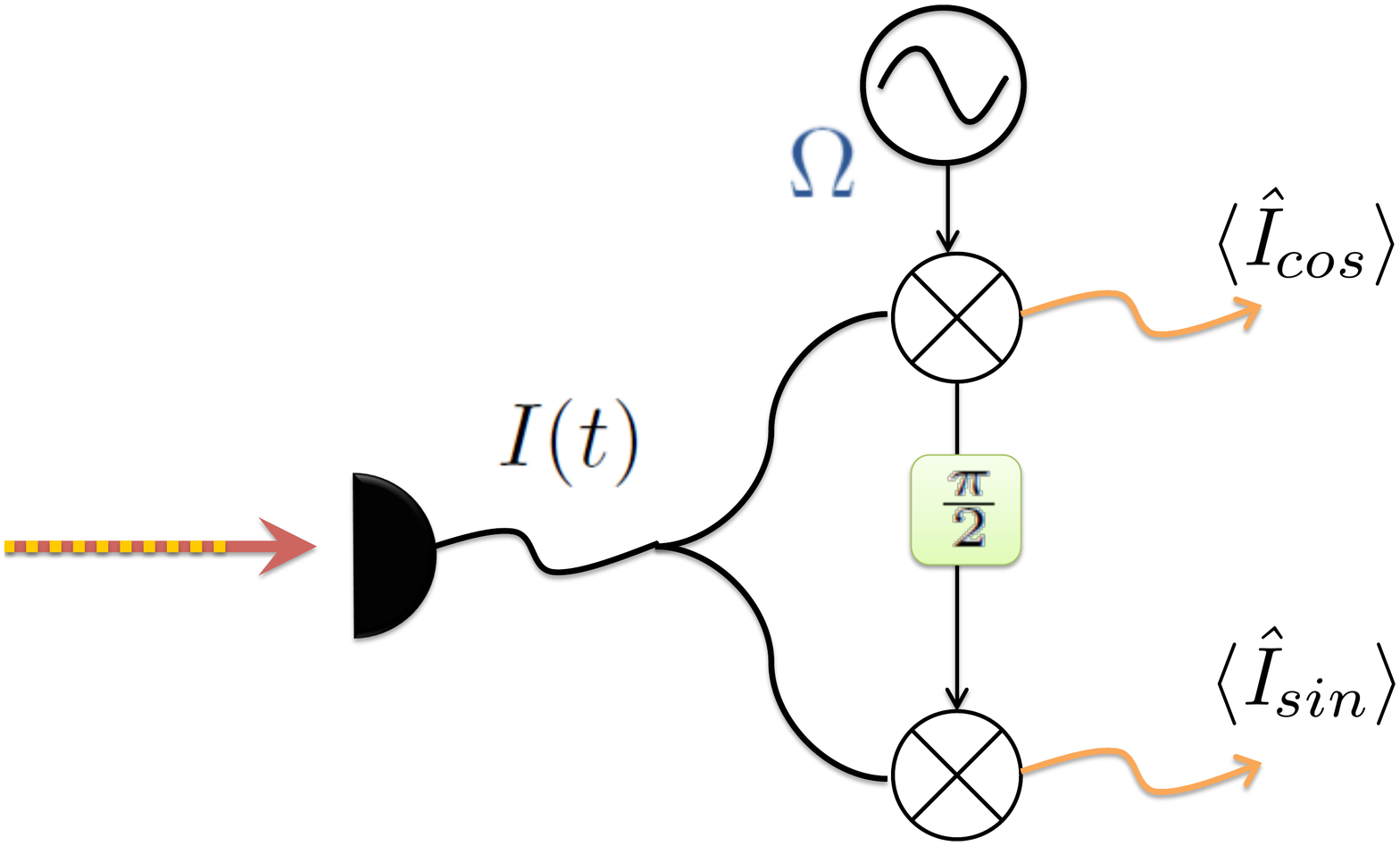}
    \caption{Photocurrent demodulation chain before acquisition board, implemented for each detector, as presented in \cite{hexaopo}.}
    \label{detect}
  \end{center}
\end{figure}

Each reflected field is divided by a balanced beam splitter (BS), and measured by a pair of photodetectors, allowing for a continued calibration of the standard quantum limit by vacuum homodyning. The output currents where demodulated by a pair of in-quadrature electronic local oscillators (Fig. \ref{detect}), set for 21 MHz, and filtered by 300 kHz low-pass filters. This results in two outputs for each photocurrent, $\hat I_\mathrm{cos}$ and $\hat I_\mathrm{sin}$. These observables are associated to the Fourier transform of the photocurrent operator as
\begin{equation}
\label{eq:Iomeganonhermitian}
\hat I_\Omega= \mathrm{e}^{-i\varphi} \hat{a}^{(i)}_{\omega_{i}+\Omega} +\mathrm{e}^{i\varphi}\hat{a}^{(i)}_{\omega_{i}-\Omega}= \hat I_\mathrm{cos}+ i \hat I_\mathrm{sin},
\end{equation}
with the phase $\varphi$ associated to the mean value of the field for each mode as $\alpha_{\omega_i}=|\alpha_{\omega_i}|\exp(i\varphi)$. These photocurrents are read by a digitalizing board as the cavities are scanned through resonance.  Therefore, combining electronic demodulation and cavity detection \cite{hexaopo, prlsideband}, we were able to reconstruct the covariance matrix of the output sidebands. Since the detected modes are of Gaussian nature \cite{gaussian}, determination of the covariance matrix is equivalent to the complete tomography of the output state of the sidebands of the intense optical fields involved.

The overall detection efficiencies are 87$\%$ for the infrared beams and 65$\%$ for the pump, accounting for detector efficiencies and losses in the beam paths.

\subsection{Entanglement Criterion}

Following~\cite{Cassemiro_2007}, quantum properties of Gaussian states are completely characterized by their second order moments, organized in the covariance matrix $\mathbf{V}=\left( \langle \vec{\mathbf{X}}\cdot \vec{\mathbf{X}}^{T} \rangle + \langle \vec{\mathbf{X}}\cdot \vec{\mathbf{X}}^{T} \rangle^{T}\right)/2$, where
\begin{equation}
\vec{\mathbf{X}}=(\hat{p}^{(1)}~ \hat{q}^{(1)}~\hat{p}^{(2)}~ \hat{q}^{(2)}\cdots ~ \hat{p}^{(N)}~ \hat{q}^{(N)})^{T},
\end{equation}
is the vector of the amplitude  [$\hat{p}^{(j)}=\hat{a}^{(j)}+\hat{a}^{(j)\dagger}$] and phase [$\hat{q}^{(j)}=-i(\hat{a}^{(j)}-\hat{a}^{(j)\dagger})$] quadrature operators and $N$ is the number of field modes. The operators $\hat{a}^{(j)}$ and $\hat{a}^{(j)\dagger}$ are the usual annihilation and creation operators for mode $j$ in any arbitrary basis. The canonical commutation relations can be written in the compact form as $[\vec{\mathbf{X}},\vec{\mathbf{X}}^{T}]=2i\mathbf{W}$, where
\begin{equation}
\mathbf{W}=\bigoplus_{j=1}^{N}\mathbf{w}, \hspace{1cm} \mathbf{w}= \left(\begin{matrix}
\phantom{-} 0 &1\\
-1 &0
\end{matrix}
\right)
\label{Omega}
\end{equation}

In order to represent a physical state, the covariance matrix must obey the Robertson-Schr\"{o}dinger uncertainty principle~\cite{Schrodinger_1930,Robertson},
\begin{equation}
\mathbf{V}+i\mathbf{W}\geq 0,
\label{inequality}
\end{equation}
which implies a condition on the symplectic eigenvalues of the covariance matrix
\begin{equation}
\nu^{(k)}\geq 1, \hspace{1cm} k=1,2,\ldots, N.
\label{eigenvalues}
\end{equation}
The symplectic eigenvalues of the covariance matrix can be obtained as the square roots of the ordinary eigenvalues of $-(\mathbf{W V})^{2}$.

A well known separability criterion relies on the positivity of the partially transposed (PPT) density matrix~\cite{Peres_1996}, that could be used to test the entanglement among all possible bipartitions of a system. This map is positive for all separable states, however it may be negative for entangled states.
For continuous variables, it is equivalent to a local inversion of time for the transposed subsystems~\cite{Simon_2000} in phase space.

The partial transposition (PT) operation over the covariance matrix turns $\hat{q}^{(n)}$ into $-\hat{q}^{(n)}$ for a determined subset of modes. If the resulting PT covariance matrix $\tilde{V}$, violates the inequality given in Eq. (\ref{inequality}), we have a sufficient condition for the existence of entanglement among the transposed subset of modes and the remaining subset~\cite{Simon_2000}, or equivalently, the symplectic eigenvalues must violate Eq. (\ref{eigenvalues}) in this case. The PPT criterion is both necessary and sufficient for pure or mixed states in partitions $1\times (N-1)$~\cite{Simon_2000,Werner_2001}. Other partitions from systems with $N\geq 2$ may possess bound entanglement, nevertheless it is always sufficient.

The smallest symplectic eigenvalue $\tilde{\nu}_{\text{min}}$ of the PT covariance matrix is useful not only to witness the entanglement but also to quantify it. In fact, the entanglement measure given by the logarithmic negativity~\cite{Vidal_2002} can be written as a decreasing function of $\tilde{\nu}_{\text{min}}$, for all $(M+N)$-mode bisymmetric Gaussian states~\cite{Adesso_2004}. Thus, a larger violation of Eq. (\ref{eigenvalues}) implies a stronger entanglement.

\subsection{Analysis of hexapartite entanglement}

The five examples presented in the main body of the article are a sample of all the 31 possible bipartitions for the six modes of the field.
 The complete results are shown in Figs. (\ref{strong}--\ref{pump}), showing the evaluation of $\tilde{\nu}_{\text{min}}$. In what follows, bipartitions are labeled by the modes described in Fig. 2 (main text) as $nl$ for the mode at frequency $\omega_n-\Omega$ (i. e., at the lower sideband), and $nu$ for modes at the upper sideband $\omega_n+\Omega$, $n$ referring to the carrier frequency. Ordering inside the brackets is from smaller to higher frequencies, assuming $\omega_1<\omega_2$.

In Fig. \ref{strong}, we see the results in configurations that are analogous to the black curve of Fig. 3 (main text). Beyond the obvious symmetry in taking upper or lower sidebands of signal and idler fields into the smallest of the  $2 \times 4$ partition, there is the curious effect that any arrangement of the pump sidebands will result in the same violation of the entanglement witness, and the same behavior for growing pump power.

\begin{figure}[h]
  \begin{center}
    \includegraphics[width= 0.95\linewidth]{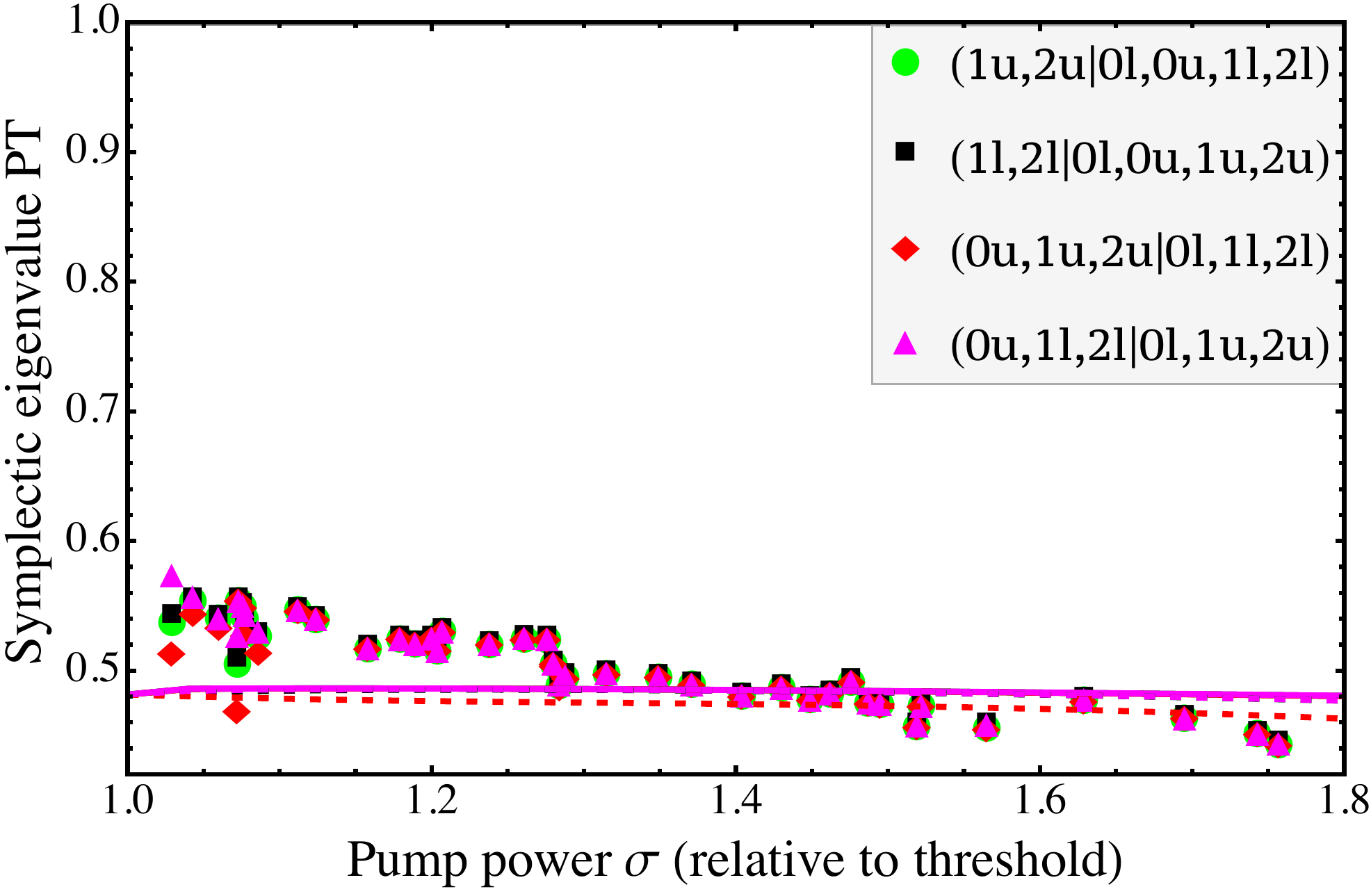}\\%{figures/ULplot15All.pdf}
    \caption{Symplectic eigenvalues from the transpositions of different bipartitions, as labeled in the inset of the graph.}
    \label{strong}
  \end{center}
\end{figure}

The situation is different if the pair of sidebands from the same beam are taken into the same bipartition, as can be seen in Fig. \ref{singlebeam}. As discussed in the analysis of the result presented in the red trace of Fig. 3 (main text), this situation is pretty sensitive to the added noise from the phonon coupling. The only diference in the presented graph is a slight increase in the violation if pump sidebands are split between the bipartitions in a $3\times3$ configuration, when compared to the $2\times 4$ bipartition.

\begin{figure}[h]
  \begin{center}
    \includegraphics[width= 0.95\linewidth]{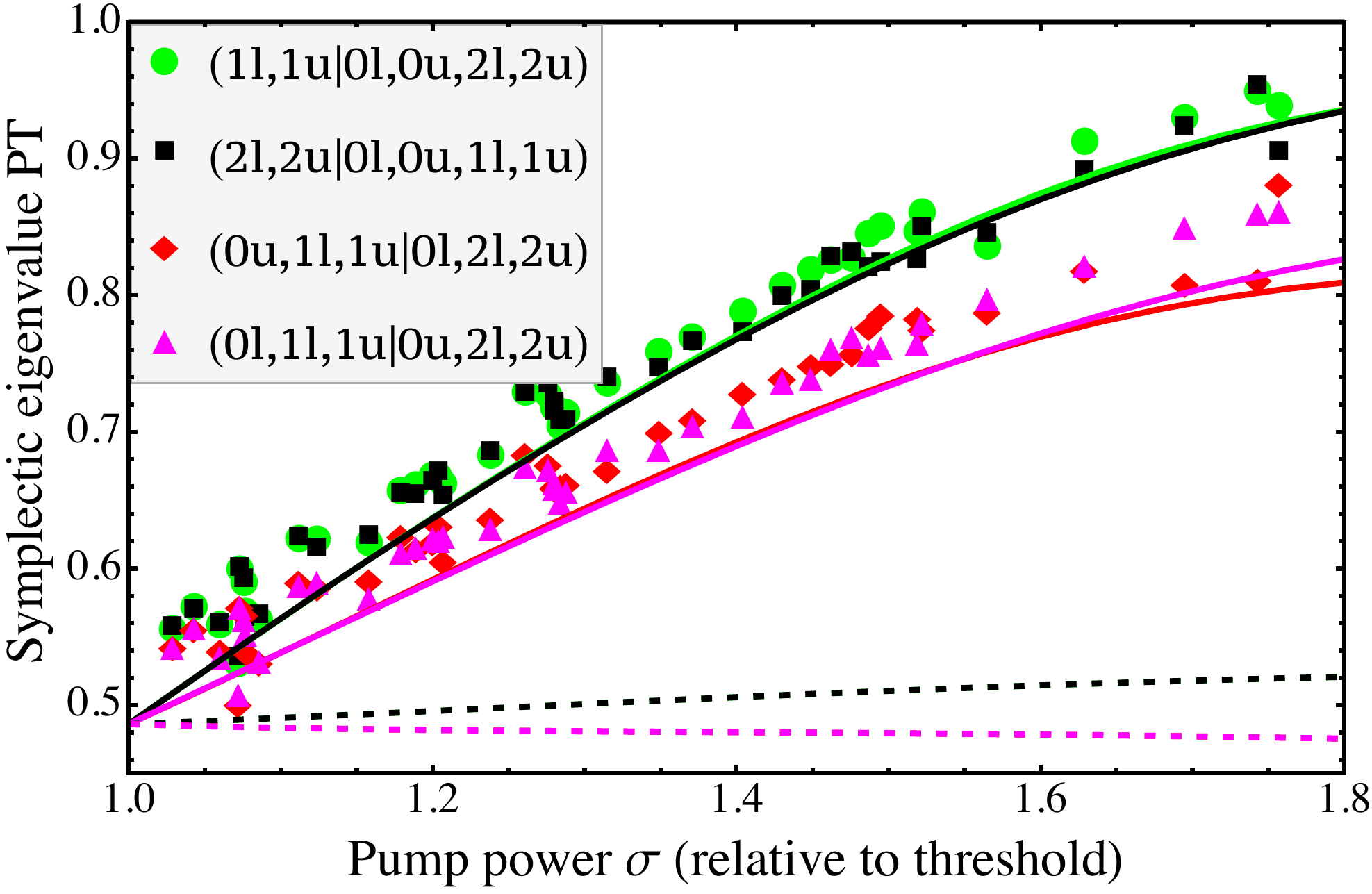}%{figures/ULplot17All.pdf}
    \caption{Symplectic eigenvalues from the transpositions of different bipartitions, as labeled in the inset of the graph.}
    \label{singlebeam}
  \end{center}
\end{figure}

\begin{figure}[h]
  \begin{center}
    \includegraphics[width= 0.9\linewidth]{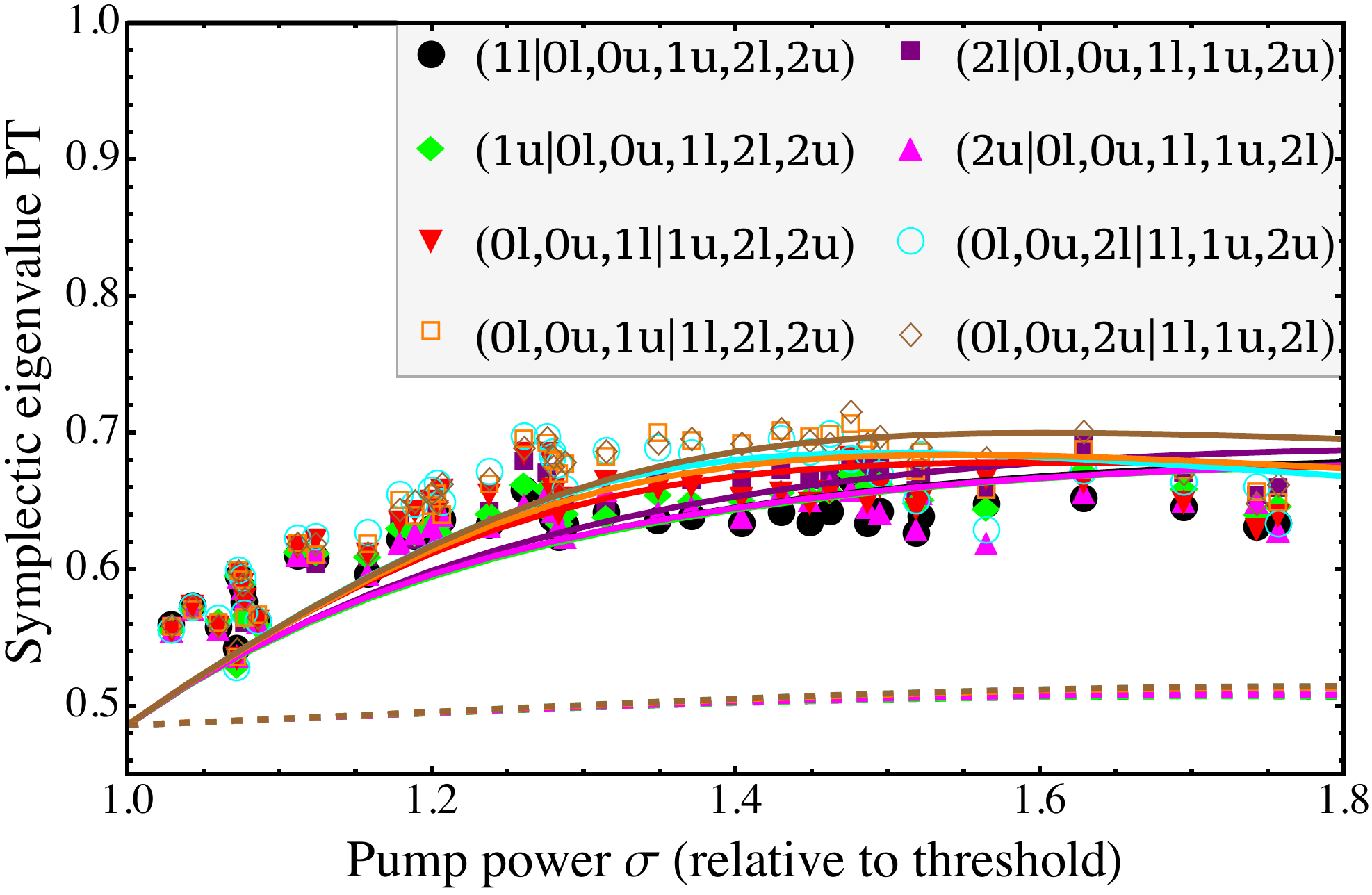}
    \includegraphics[width= 0.9\linewidth]{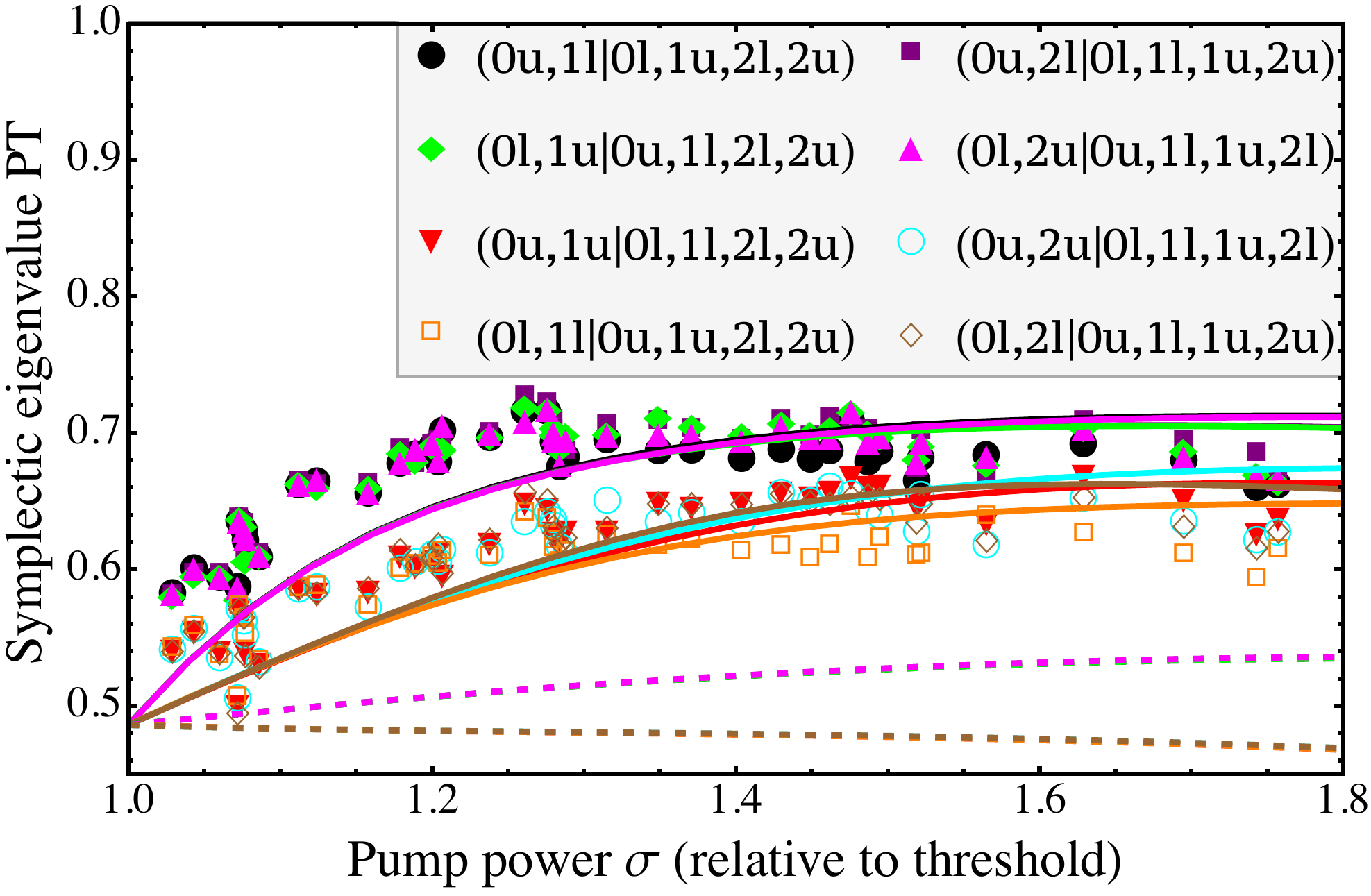}
    \caption{Symplectic eigenvalues from the transpositions of different bipartitions, as labeled in the inset of the graph.}
    \label{onemode}
  \end{center}
\end{figure}

It is curious to notice that if we take a single sideband of the downconverted fields, the symplectic eigenvalue is smaller than that with the contribution of both sidebands, as seen in the blue curve of Fig. 3 (main text). All the 16 possible configurations show similar results (Fig. \ref{onemode}), with a slightly stronger violation for $2\times 4$ bipartitions where just one of pump sidebands, that one linked to the downconverted mode by the beam splitter operation, is taken in the same set.

\begin{figure}[h]
  \begin{center}
\includegraphics[width= 0.95\linewidth]{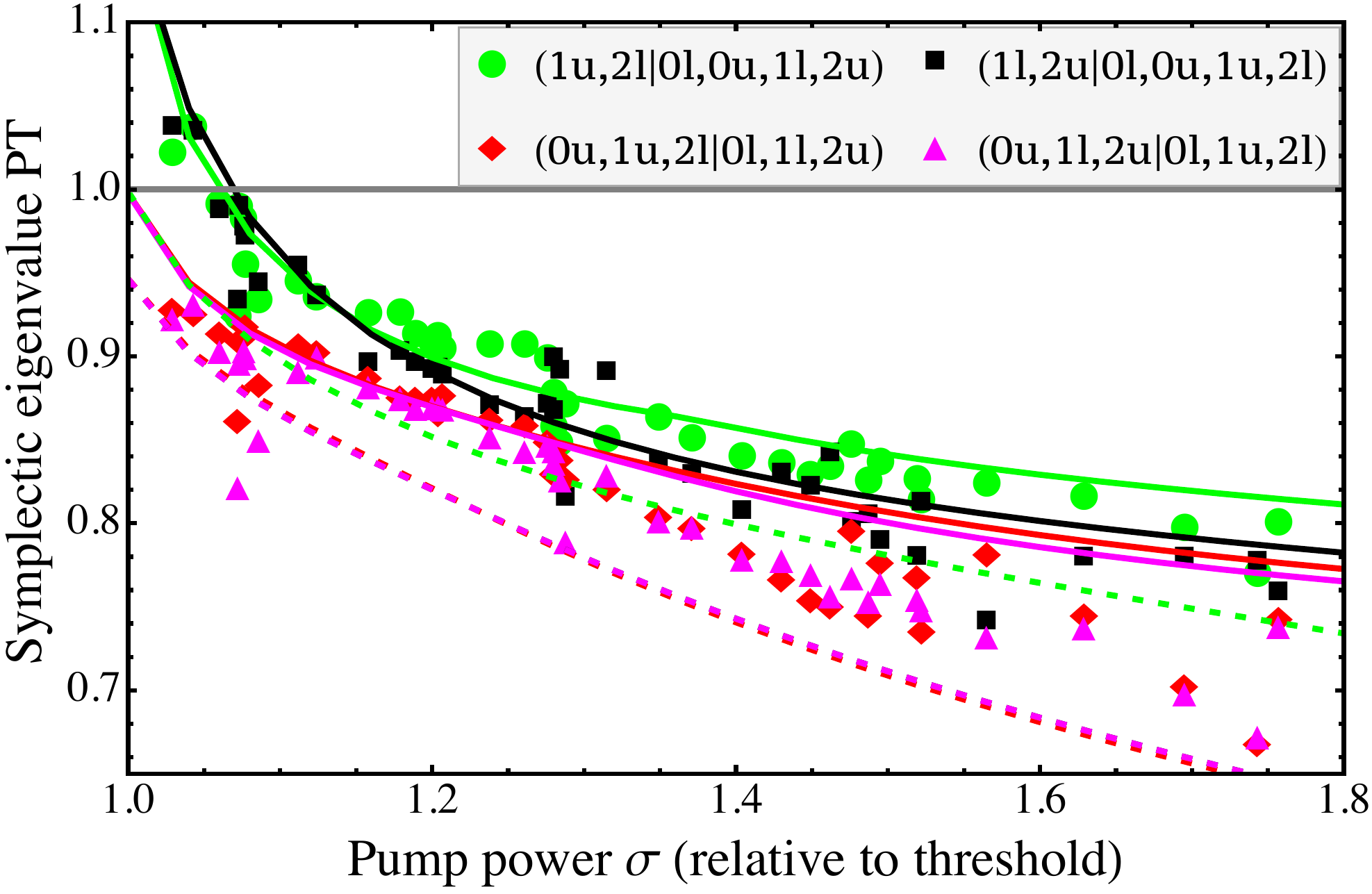}\\%{figures/ULplot18All.pdf}
    \caption{Symplectic eigenvalues from the transpositions of different bipartitions, as labeled in the inset of the graph.}
    \label{squeezer}
  \end{center}
\end{figure}

All those configurations involved splitting of modes linked by the two mode squeezing operators, and evidently begin at a maximized violation at the oscillation threshold. The coupling between the pair of entangled modes ($1l2u \times 1u2l$) is negligible close to threshold, but through the beam splitter operation, both pairs of modes are coupled to the sidebands of the pump. This coupling, proportional to signal and idler mean fields, increases with the pump power and leads to a growing entanglement between these fields, as can be seen in Fig. \ref{squeezer}. From this set of data we have selected the case presented in Fig. 4 (black) in the main text. There is a small gain if the pump sidebands are split between the subsets. Phonon noise affects entanglement, yet in a less dramatic way as in the other situations presentes in Figs. \ref{singlebeam},\ref{onemode}, and \ref{pump}.

The last possible combination involves the entanglement of the pump sidebands against the rest of the system. This situation is directly related to the tripartite entanglement previously verified \cite{tripartite}.
The three possibilities are presented in Fig. \ref{pump}. Inseparability of the pump sidebands from the downconverted modes is presented also in Fig. 4 (red) of the main text, and shown here in comparison with the separability of each sideband. This partition is the most affected by phonon noise, since entanglement should grow for increasing pump powers, but the added noise increases in a faster pace, when compared to the dynamics of Fig. \ref{squeezer}. Therefore, the pump sidebands act as a common reservoir for the two mode squeezed states, leading to their entanglement, and becoming entangled as well. On the other hand, since each pair of modes of a single beam is coupled to a thermal bath, correlation is degraded for growing coupling.

\begin{figure}[h]
  \begin{center}
    \includegraphics[width= 0.95\linewidth]{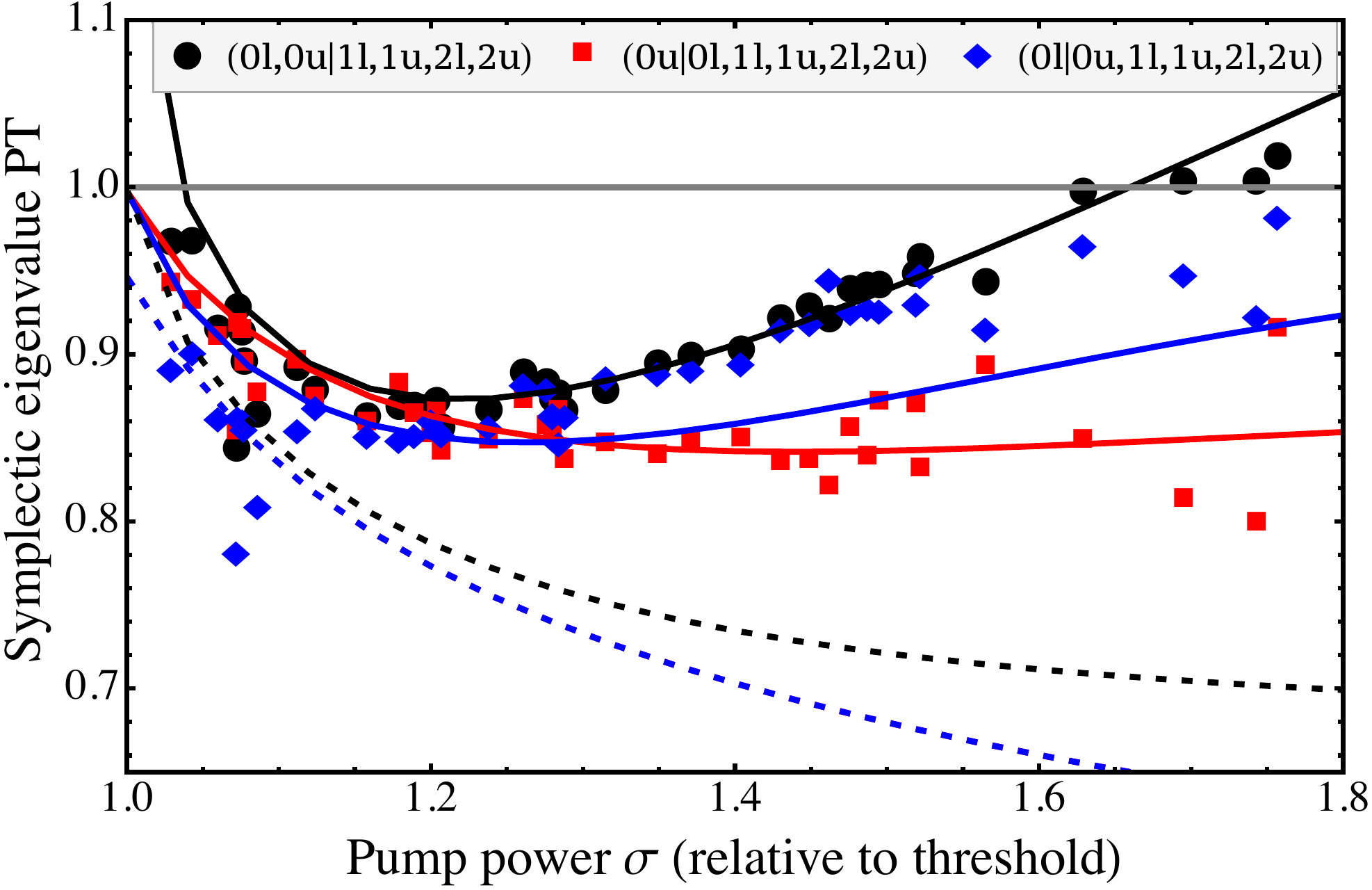}%{figures/ULplot18All.pdf}
    \caption{Symplectic eigenvalues from the transpositions of different bipartitions, as labeled in the inset of the graph.}
    \label{pump}
  \end{center}
\end{figure}

One word should be said about the uneven role of upper and lower sidebands in Fig. (\ref{pump}). This comes from the power imbalance of the sidebands observed in the system \cite{Munos}. A simple cavity model would consider perfect resonance for the carriers, but in real experimental conditions, some cavity detuning will arise from imperfect locking, leading to the coupling of amplitude and phase quadratures inside the cavity. We have a careful locking of the signal and idler modes, but pump detuning is adjusted by temperature, that may drift by optical feedback coming from the absorption of the crystal. As a result, the sidebands of the pump field may be unevenly coupled, and entanglement may become asymmetric, as observed in Fig. \ref{pump}. If the theory allows for some detuning, the result presents a good agreement with experimental data.

\end{document}